# Android Malware Detection Using Autoencoder

Abdelmonim Naway[1], Yuancheng LI[1]

[1] North China Electric Power University, School of Control and Computer Engineering, 2 Beinong Road, Changping District, Beijing, China,102206 {abdelmonim, yuanchengli}@ncepu.edu.cn

**ABSTRACT:**
*Smartphones have become an intrinsic part of human's life. The smartphone unifies diverse advanced characteristics. It enables users to store various data such as photos, health data, credential bank data, and personal information. The Android operating system is the prevalent mobile operating system and, in the meantime, the most targeted operating system by malware developers. Recently the unparalleled development of Android malware put pressure on researchers to propose effective methods to suppress the spread of the malware. In this paper, we propose a deep learning approach for Android malware detection. The proposed approach investigates five different feature sets and applies Autoencoder to identify malware. The experimental results show that the proposed approach can identify malware with high accuracy.*

**Keywords:** Android Malware Detection; Deep Learning, Autoencoder; Static Malware Analysis

## [1] INTRODUCTION

The distribution of Android powered devices is estimated to be 1.2 billion in 2018. The forecast of global distributions in 2022 indicates that there will be 1.4 billion Android Devices [26]. Furthermore, the number of obtainable Android applications from the official Google play store is reached 3 million Android apps in 2017. The present rate of development is more than 1,300 apps a day [27]. The wide use of the Android operating system and proliferation of apps concurred with an unprecedented number of Android malware that reach 9,411 sophisticated malware daily for the Android in 2018. This means new malware emerging every 10 seconds [28].

In an effort to defend Android users and their valuable information, many malware detection techniques have been proposed. Conventional methods such as signature based and heuristic based identification of antivirus can only distinguish already identified malware and consequently limit their detection effectiveness. Recently, deep learning has been introduced for Android malware detection. Deep learning is an explicit subfield of machine learning: a new method for learning representations from data that focus on learning consecutive layers of progressively significant representations (deep in deep learning refers to the concept of consecutive layers of representations. How many layers are devoted to a model of the data is known as the depth of the model) [24].

In light of this background, we propose a method for Android malware detection, employing static analysis and depend on five features extracted from APK files to build a deep learning classifier based on Autoencoders to classify Android apps into benign and malicious. In summary, our main contributions are twofold: first, developing a new Android malware detection system based on deep learning. Second, empirical evaluation of the proposed approach on a recent real-world dataset that reflects the changes in features according to the changes in Android specifications and contains new malware types.

The rest of the paper is organized as follows: Section 2 introduces the concerned work on Android malware detection and deep learning techniques. Section 3 describes the architecture of the proposed methodology. The experimental results are illustrated in section 4 and the conclusion follows in section 5.

# [2] Related Studies

## [2-1] Android Malware Detection Using Static Analysis

The static analysis screen parts of the application without really executing them. W. Li et al. [15] designed a malware identification system based on a deep belief network. They suggested two types of features from Android apps for malware characterization, namely risky permissions and API function calls. R. Nix et al. [16] they focused on program analysis that observes Android API calls made by an application. System API calls describe how an application communicates information with the Android OS. Such communication is crucial for an application to do its jobs, consequently providing essential data on an application's behavior. K. Xu et al. [17] developed the DeepRefiner malware identification system employs deep neural networks with diverse hidden layers. In a preprocessing step, DeepRefiner retrieves XML values from XML files in the first detection layer and grabs bytecode semantics from the disassembled classes.dex file in the second detection layer. DeepRefiner then connotes apps as vectors, which are used as inputs for deep neural networks. The hidden layers in neural systems accordingly build identification features from input vectors through non-linear translation.

W. Wang et al. [18] seeking to improve the accuracy of large-scale Android malware detection by developing a hybrid system based on deep autoencoder (DAE) as pre-training procedure and different CNN structures for malware identification. Experimental results showed that CNN-P structure accomplished the best accuracy. Yi Zhang et al. [19] They proposed an identification approach stand on Convolutional Neural Network (CNN) and implemented a system DeepClassifyDroid. The structural design of DeepClassifyDroid is composed of three components: feature extraction component; embedding in vector space (embedding different feature sets into a joint vector space); deep learning model uses convolutional neural networks to carry out malware characterization. Dali Zhu et al. [20] presented DeepFlow, a malware detection system that established on data streams within malware apps that may differ essentially from ones within benign apps, however, might be like different malignant apps to some degree. DeepFlow uses such differences and similarities to consequently determine new apps whether they are malignant or not by utilizing a deep learning model.

## [2-2] Android Malware Detection Using Dynamic Analysis

The dynamic analysis technique includes the execution of the application on either a virtual machine or a physical device. H. Liang et al. [21] developed natural language processing techniques for Android malware analysis on the premise that there is a resemblance between theme drawing and malware identification. They designed a model that deals with system call sequences as texts and considers the malware detection function as theme extraction. First, an embedding layer was utilized to represent the system calls to vector space. Then, the module, which is called vertical multiple convolutions, was employed to exploit possible high-level information from the exceeding presentation matrices of the long sequence. Finally, a multilayer perceptron with SoftMax layer was applied to complete the classification function.

## [2-3] Android Malware Detection Using Hybrid Analysis

The hybrid analysis technique consolidates static analysis and dynamic analysis features. R. Vinayakumar et al. [22] proposed the use of Long Short-Term Memory (LSTM) which is a special type of recurrent neural network used to study long-term transient dynamics with a series of random lengths for Android malware identification based on static and dynamic features. The results showed a good performance accomplished by LSTM. H. Alshahrani et al. [23] developed DDefender a malware identification system includes two main parts: first, client side, a light app running on the user's phone to preform dynamic analysis and provide analysis report for the users. Second, server side, a system that preforms static analysis and identification procedure and sends the outcomes back to the client side.

In this paper, we intend to use the Autoencoder in a different way from the way it has been used in literature. Usually, Autoencoder is used for features reduction or used in a pre-training phase, then another algorithm is applied for classification. Here we will use the Autoencoder to classify Android apps into benign or malicious ones. This requires training Autoencoder in semi - supervised way. The trained model will be tested on the labeled and unlabeled dataset.

## [3] The proposed Methodology

[Figure-1] demonstrate the proposed methodology. The proposed methodology includes the following steps:

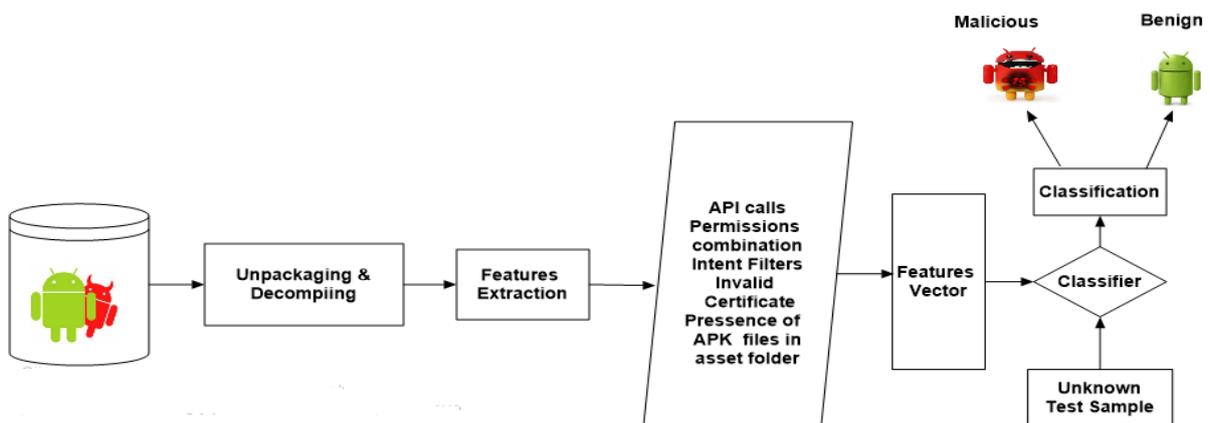

Figure 1: The proposed methodology

### [3-1] Apps de-compilation

The Mobile Security Framework (MobSF) [16] which, is a tool for Android malware analysis that can perform static analysis and dynamic analysis is used to decompile the apps and generation of smali files.

### [3-2] Features Extraction

The MobSF is utilized for features extraction first, from Androidmanifest.xml, which include the Permissions and Intent filters. Then the APIs calls are drawn from Smali files created by MobSF. Afterword asset folder is checked for the presence of APK files. Finally, the legitimacy of the certificate is validated.

### [3-3] Features set

- Permissions combination(featureset1-fs1): A permission is a limit restraining the entry to a portion of the code or data on a device. The limitation is imposed to safeguard essential information and code that could be misused to damage the user's experience. Permissions are utilized to allow or limit an application access to restricted APIs and resources [3]. Permissions that are asked for by an application can imply its functionality to some extent. Nevertheless, permissions can serve both malicious apps and benign ones. Thereby, it is necessary to join various permission requests concurrently for proper malware identification [4], [5].
- Intent Filters(featureset2-fs2): Intent is an intricate messaging system in the Android system, and is conceived as a security technique to block applications from obtaining passage to other applications instantly. Applications are required to have particular permissions to utilize Intents. This is a method for regulating what applications can do once they are set up in Android. Intent-filter – characterized in AndroidManifest.xml file – reports the kind of Intent the application is able of taking [6], [7].
- API Calls(featureset3-fs3): APIs utilized by an application chooses the genuine convenience and limit of the application. Static analysis of APIs handled in an application is pivotal to relate to what the application truly plans to do. Particular API calls enable access to system services or resources of the device and are frequently found in malicious applications. As these calls can be especially direct to malignant conduct [8]. The following API calls are analyzed: Telephony Manager. [9]; HTTP association and attachments [10]; DexcClassLoader [11]; Reflection; System Service; Runtime and System; cryptographic activities (crypto) [12].
- Invalid certificate(featureset4-fs4): To examine the legality of the certificate. An invalid certificate demonstrates that the application has tampered with (repackaged) [13].
- Presence of APK files in the asset folder(featureset5-fs5): masking data in the asset folder is regarded as an exemplary behavior of the malicious application. [14]

**[3-4] Feature Vector**

All of the extracted features are converted into a feature vector. Each feature set is described as a Boolean expression with different dimensions and then produced a unified representation by mapping different sets into a joint vector space. For evaluation, the defined combination of defined feature set S:

$$S = fs1 \cup fs2 \cup fs3 \cup fs4 \cup fs5$$

| $S_i$ |- dimension set $S_i$ is used to represent a feature set which is a vector of zeros with a '1' in the position if apps have used a certain feature in the feature set. Consequently, we can convert any app X to a vector space $\varnothing(x)$:

$$\varnothing : x \rightarrow \{0,1\}^{|s|}, \varnothing(x) \rightarrow (I(x,s))s \in S$$

Where the indicator function $I(x,s)$ is defined as:

$$I(x,s) = \begin{cases} 1 & \text{if the application } x \text{ contains feature } s \\ 0 & \text{otherwise} \end{cases}$$

Thus, different features can be mapped to into unified joint vector.

**[3-5] Deep learning Classifier**

Building upon the extracted features the Autoencoder study features to identify Android apps. The Autoencoder as illustrated in **[Figure- 2]** below comprises of an encoder and a decoder. The correspondence of the input layer to the hidden layer is known as encoding and the correspondence of the hidden layer to the output layer is known as decoding. The encoder receives the vector of input characteristics and translates them via sigmoid activation functions in the hidden layers into new features [1].

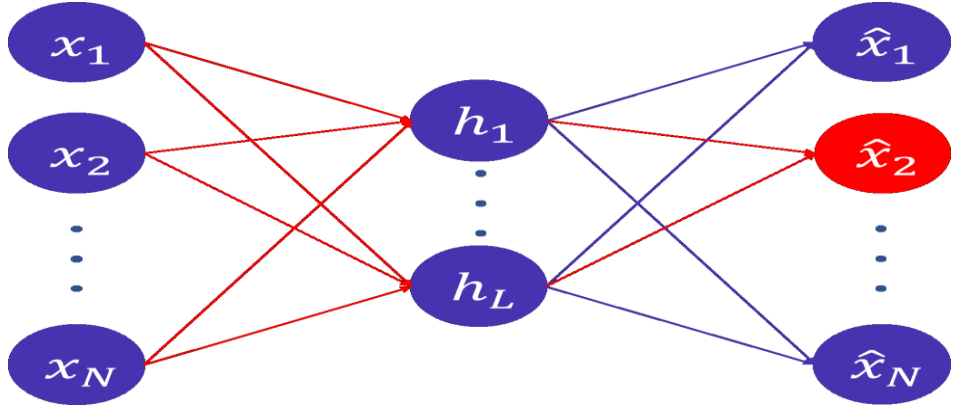

**Figure 2: Autoencoder Structure**

An autoencoder is a feed forward neural network utilized to study representations of data. The principle is to prepare a network with at least one hidden layer to rebuild its inputs. The output values are established equal to input values, i.e. $\hat{x}$ = x for the purpose of studying the identity transformation h(x) ≈ x. The autoencoder accomplishes this by arranging input characteristic to hidden layer nodes utilizing an encoder function [1]:

$$h(x) = f(Wx + b_h) \quad (1)$$

where x is the input vector of characteristics, f stands for the sigmoid function, $b_h$ is the vector of hidden neuron biases, and W is the matrix of hidden weights. The data is restored utilizing a linear decoder:

$$g(\hat{x}) = \lambda f(Wx + b_h) + b_g \quad (2)$$

After studying the weights W, every hidden neuron depicts a specific feature of the input data. Therefore, the hidden layer h(x) can be regarded as a new feature depiction of the input data. The hidden depiction h(x) is then utilized to rebuild an estimate $\hat{x}$ of the input utilizing the decoder function g($\hat{x}$)[1]. The frequently used function for the encoder and decoder is a nonlinear sigmoid function. Training is generally accomplished by decreasing the squared reconstruction error $(g(\hat{x}) - x)^2$ using the backpropagation by gradient descent algorithm. in practical terms, the restriction $\lambda = W^T$ is usually determined to lower the number of parameters [1]. The fundamental idea is that by learning the hidden depiction h(x) that can rebuild the original input characteristics, the autoencoder obtains significant features of the input characteristics.

The training procedure for an autoencoder can be separated into two phases: the first phase is to learn features and the second phase is to adjust the network. In the first phase, feed-forward propagation is first executed for each input to get the output value $\hat{x}$. Then the errors are utilized to quantify the deviation of $\hat{x}$ from the input value. In the end, the error will be backpropagated through the network to revise the weights. In the adjusting phase, with the network having appropriate features at each layer, the standard supervised learning method can be adopted to tweak the parameters at each layer [2].

**[3-6] Classification**

Classification is the procedure for studying the target function $f$ that interprets each attribute set $x$ to one of the predetermined class labels $y$. By applying the Autoencoder, the new unknown apps will be labeled as benign or malicious.

## [4] Result and Discussion

### [4-1] Dataset Setup

For all experiments, we use a dataset of real Android apps and malware. The dataset contains 1200 apps in total, incorporating 600 benign apps and 600 malicious apps. The benign apps were downloaded from the official Google play store (https://play.google.com/store) and APKpure (https://apkpure.com) in September 2017. The benign apps belong to different categories such as magazines, utilities, games, etc.., and represent popular apps in their corresponding category (all the downloaded apps have at least 500,000 installations). All of the benign apps passed the scan of the latest versions of 3 Antivirus, ESET Internet Security 11.2, Symantec Endpoint 14, and Kaspersky Internet Security 18. The malicious apps obtained from Contagiodump(http://contagiominidump.blogspot.com), droidbench (https://github.com/secure-software-engineering/DroidBench), Android malware (github.com/ashishb/android-malware), virusshare (https://virusshare.com), and virussign (www.virussign.com) represent different malware types such as Trojan horses, Backdoors, Information Stealers, Ransomware, Scareware, etc... The malware was picked from datasets collected between May 2013 and December 2017.

### [4-2] Experimental Setup

The experiments are conducted using Autoencoder model utilizing Keras library. **[Table- 1]** below shows the structure of the Autoencoder.

Table 1: Autoencoder Parameters

```
Layer (type)                 Output Shape              Param #
=================================================================
input_1 (InputLayer)         (None, 40)                0
_________________________________________________________________
dense_1 (Dense)              (None, 200)               8200
_________________________________________________________________
dense_2 (Dense)              (None, 100)               20100
_________________________________________________________________
dense_3 (Dense)              (None, 100)               10100
_________________________________________________________________
dense_4 (Dense)              (None, 40)                4040
=================================================================
Total params: 42,440
Trainable params: 42,440
Non-trainable params: 0
_________________________________________________________________
```

As shown in **[Table-1]** the Autoencoder uses 4 fully connected layers. The first two layers are used for the encoder, the last two layers go for the decoder. The sigmoid activation function is used in the first and fourth layers, while relu and tanh activation functions are used on the second and third layers respectively. Autoencoder trained on malicious apps only. Keeping the benign apps on the test set will provide a means to evaluate the performance of the Autoencoder. 80% of the dataset is used for training, while 20% of the dataset is kept for testing. We chose accuracy and F1-Score to evaluate the experiments.

**[4-3] Experimental Result**

Table 2: The Autoencoder detection result

| ACCURACY | F1-SCORE |
|---|---|
| 96.81 | 95.4 |

As shown in **[Table-2]** the ability of the Autoencoder to identify an app as benign or malicious is 96.81%. Further, the F1-score which indicate to what extent the Autoencoder model determinant is 95.4%.

**[Figure-3]** illustrates the loss in the training and test datasets. **[Figure- 4]** illustrates the accuracy in training and test datasets.

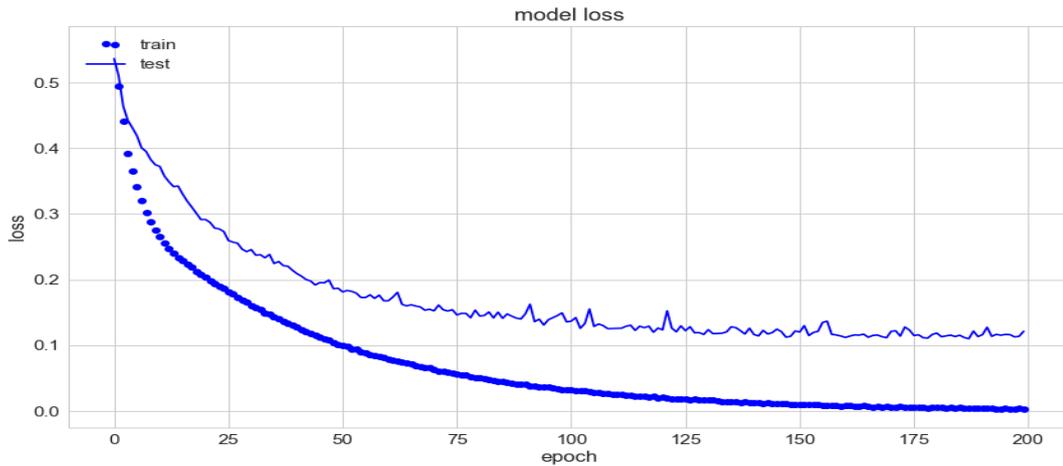

Figure 3: The loss in training and test datasets

As demonstrated in **[Figure-3]** the reconstruction error in training and test data appears to converge well, and it is small enough.

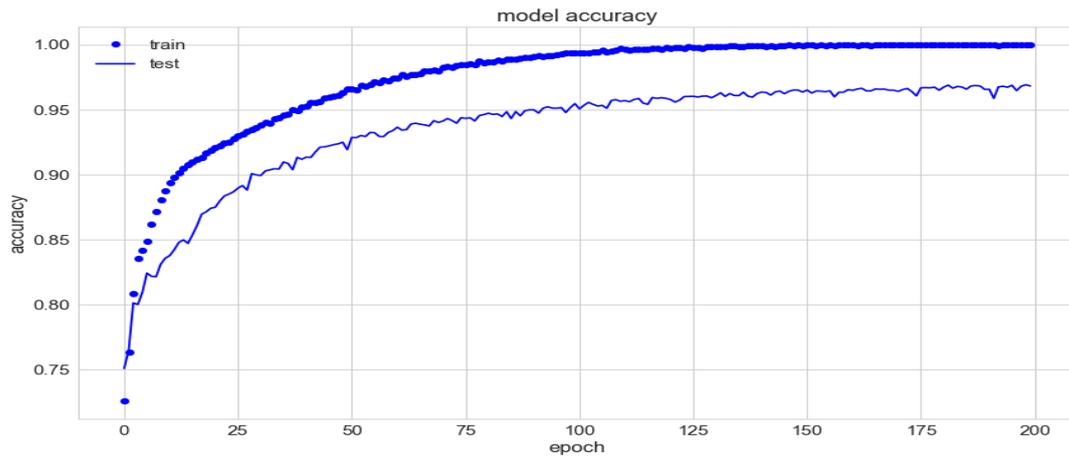

Figure 4: Accuracy in training and test datasets

As demonstrated in **[Figure- 4]** the accuracy in training and test data is high and on average, the Autoencoder model achieved 96.81% of accuracy.

For further evaluation of the model, we conducted several experiments using different dataset split as demonstrated in **[Table- 3]**.

**Table 3:** Demonstrates Accuracy and F1-Score using a different split for training and test dataset

| *Train test split* | *Accuracy* | *F1-Score* |
|---|---|---|
| 70%-30% | 96.39 | 94.79 |
| 60%-40% | 96.57 | 95.07 |
| 50%-50% | 94.81 | 92.38 |

As shown in Table 3 the detection performance changes with the Autoencoder of different dataset split percentage but remained high.

# [5] Conclusion

The surge of Android malware that appears every 10 seconds as reported in [28] call for a new malware detection technique that can live up to the damage the malware can inflict on Android powered devices. In this paper, we propose a new, deep learning-based approach to Android malware detection resting on five different features: permissions combination; Intent filters; API calls; Invalid certificates; the presence of APK files in asset folder to construct an Autoencoder classification model that is capable of identifying malicious apps from benign ones. The performance of the proposed approach is evaluated on real-world dataset contain different benign apps categories and various malicious sample types. The result shows that the proposed approach can achieve high accuracy of 96.81%.

REFERENCES
[1]. Lewis, N. "The Secret to the Autoencoder " Deep Learning Made Easy with R A Gentle Introduction to Data Science (2016), 119-121
[2]. Liu, Weibo, et al. "A survey of deep neural network architectures and their applications." Neurocomputing 234 (2017): 11-26.
[3]. Tchakounté, F., and F. Hayata. "Supervised Learning Based Detection of Malware on Android." Mobile Security and Privacy. 2016. 101-154.
[4]. S. Liang and X. Du, "Permission-combination-based scheme for Android mobile malware detection," *2014 IEEE Int. Conf. Common. ICC 2014*, pp. 2301–2306, 2014.
[5]. W. Li, J. Ge, and G. Dai, "Detecting Malware for Android Platform: An SVM-Based Approach," *Proc. - 2nd IEEE Int. Conf. Cyber Secur. Cloud Compute. CSCloud 2015 - IEEE Int. Symp. Smart Cloud, IEEE SSC 2015*, pp. 464–469, 2016.
[6]. F. Idrees, M. Nagarajan, M. Conti, T. M. Chen, and Y. Rahulamathavan, "PIndroid: A novel Android malware detection system using ensemble learning methods," *Compute. Secur.*, vol. 68, pp. 36–46, 2017.
[7]. A. Feizollah, N. B. Anuar, R. Salleh, G. Suarez-Tangil, and S. Furnell, "AndroDialysis: Analysis of Android Intent Effectiveness in Malware Detection," Compute. Secur., vol. 65, pp. 121–134, 2017.
[8]. D. Arp, M. Spreitzenbarth, M. Hübner, H. Gascon, and K. Rieck, "Drebin: Effective and Explainable Detection of Android Malware in Your Pocket," in Proceedings 2014 Network and Distributed System Security Symposium, 2014.
[9]. S. Y. Yerima, S. Sezer, and I. Muttik, "Android Malware Detection Using Parallel Machine Learning Classifiers," 2014 Eighth Int. Conf. Next Gener. Mob. Apps, Serv. Technol., pp. 37–42, 2014.
[10]. S. H. Seo, A. Gupta, A. M. Sallam, E. Bertino, and K. Yim, "Detecting mobile malware threats to homeland security through static analysis," *J. Netw. Compute. Appl.*, vol. 38, no. 1, pp. 43–53, 2014.
[11]. G. McWilliams, S. Sezer, and S. Y. Yerima, "Analysis of Bayesian classification-based approaches for Android malware detection," IET Inf. Secur., vol. 8, no. 1, pp. 25–36, 2014.
[12]. Aafer, Yousra, Wenliang Du, and Heng Yin. "Droidapiminer: Mining api-level features for robust malware detection in android." International conference on security and privacy in communication systems. Springer, Cham, 2013.

[13]. G. Suarez-Tangil, S. K. Dash, M. Ahmadi, J. Kinder, G. Giacinto, and L. Cavallaro, "DroidSieve," in Proceedings of the Seventh ACM on Conference on Data and Application Security and Privacy - CODASPY '17, 2017, pp. 309–320.
[14]. Hahn, Sebastian, Mykola Protsenko, and Tilo Müller. "Comparative evaluation of machine learning-based malware detection on android." Sicherheit 2016-Sicherheit, Schutz und Zuverlässigkeit (2016).
[15]. W. Li, Z. Wang, J. Cai, and S. Cheng, "An Android Malware Detection Approach Using Weight-Adjusted Deep Learning," 2018 Int. Conf. Comput. Netw. Commun., pp. 437–441, 2018.
[16]. R. Nix and J. Zhang, "Classification of Android apps and malware using deep neural networks," 2017 Int. Jt. Conf. Neural Networks, pp. 1871–1878, 2017.
[17]. K. Xu, Y. Li, R. H. Deng, and K. Chen, "DeepRefiner: Multi-layer Android Malware Detection System Applying Deep Neural Networks," 2018 IEEE Eur. Symp. Secur. Priv., pp. 473–487, 2018.
[18]. W. Wang, M. Zhao, and J. Wang, "Effective Android malware detection with a hybrid model based on deep autoencoder and convolutional neural network," J. Ambient Intell. Humaniz. Comput., vol. 0, no. 0, pp. 1–9, 2018.
[19]. Zhang, Yi, Yuexiang Yang, and Xiaolei Wang. "A Novel Android Malware Detection Approach Based on Convolutional Neural Network." Proceedings of the 2nd International Conference on Cryptography, Security, and Privacy. ACM, 2018
[20]. Zhu, Dali, et al. "DeepFlow: Deep learning-based malware detection by mining Android application for abnormal usage of sensitive data." Computers and Communications (ISCC), 2017 IEEE Symposium on. IEEE, 2017.
[21]. H. Liang, Y. Song, and D. Xiao, "An end-To-end model for Android malware detection," 2017 IEEE Int. Conf. Intell. Secur. Informatics Secur. Big Data, ISI 2017, pp. 140–142, 2017.
[22]. R. Vinayakumar, K. P. Soman, P. Poornachandran, and S. Sachin Kumar, "Detecting Android malware using Long Short-term Memory (LSTM)," J. Intell. Fuzzy Syst., vol. 34, no. 3, pp. 1277–1288, 2018.
[23]. H. Alshahrani, H. Mansourt, S. Thorn, A. Alshehri, A. Alzahrani, and H. Fu, "DDefender: Android application threat detection using static and dynamic analysis," 2018 IEEE Int. Conf. Consum. Electron., pp. 1–6, 2018.
[24]. Chollet, François, and Joseph J. Allaire. "what is deep learning?" Deep Learning with R. Manning Publications Company, 2018. 4-6.
[25]. Dalziel, Henry, and Ajin Abraham. Automated Security Analysis of Android and IOS Applications with Mobile Security Framework. Syngress, 2016
[26]. I. D. C. (IDC), "Smartphone Market Share." [Online]. Available: https://www.idc.com/promo/smartphone-market-share/os
[27]. Statista, "Statista an online statistic, market research, and business intelligence portal." [Online]. Available: https://www.statista.com/statistics/266210/number-of-available-applications-in-the-google-play-store/.
[28]. G. Data Software, "New malware every 10 seconds." [Online]. Available: https://www.gdatasoftware.com/blog/2018/05/30735-new-malware-every-10-seconds.